\documentclass[aps,prl,twocolumn,showpacs,preprintnumbers,superscriptaddress,amsmath,amssymb]{revtex4}

\usepackage{graphicx}
\usepackage{dcolumn}
\usepackage{bm}

\begin{document}
\title{Scalable fault-tolerant quantum computation in DFS blocks}
\author{Zheng-Wei Zhou}
\affiliation{Key Laboratory of Quantum Information, University of
Science and Technology of China, Hefei 230026, P.R.China}
\author{Bo Yu}
\affiliation{Key Laboratory of Quantum Information, University of
Science and Technology of China, Hefei 230026, P.R.China}
\author{Xingxiang Zhou}
\affiliation{Superconducting Digital Electronics Laboratory,
Electrical and Computer Engineering Department, University of
Rochester,\\ Rochester, New York 14627}
\author{Marc J. Feldman}
\affiliation{Superconducting Digital Electronics Laboratory,
Electrical and Computer Engineering Department, University of
Rochester,\\ Rochester, New York 14627}
\author{Guang-Can Guo}
\affiliation{Key Laboratory of Quantum Information, University of
Science and Technology of China, Hefei 230026, P.R.China}

\date{\today}

\begin{abstract}
We investigate how to concatenate different decoherence-free
subspaces (DFSs) to realize scalable universal fault-tolerant
quantum computation. Based on tunable $XXZ$ interactions, we
present an architecture for scalable quantum computers which can
fault-tolerantly perform universal quantum computation by
manipulating only single type of parameter. By using the concept
of interaction-free subspaces we eliminate the need to tune the
couplings between logical qubits, which further reduces the
technical difficulties for implementing quantum computation.
\end{abstract}

\maketitle

The fragility of quantum superposition is the major stumbling
block to achieving the physical implementation of quantum
computers. As a potential approach to prevent decoherence, DFS
encoding of qubits shields quantum coherence from environmental
noise\cite{DFS,DFSsys}. The essence of this noise immune proposal
is the assumption that the quantum systems suffer identical
environmental noise. At present, under the assumption of
collective decoherence the theory of universal fault-tolerant
quantum computation (QC) in the DFSs has been established \cite
{UFTQC1,UFTQC1a,UFTQC2}. However, the assumption of collective
decoherence requires that all qubits reside within one wavelength
of the environmental noise. This directly places a severe limit on
the scalability of QC in DFS.

To overcome this difficulty, a natural strategy is to partition the whole
qubit system into several DFS subblocks. Coherent quantum information can be
stored in them. To realize scalable fault-tolerant quantum computation
(SFTQC) in DFSs we must be able to implement local universal fault-tolerant
operations in every DFS subblock and realize fault-tolerant entangling
operations between adjacent subblocks. Recently, several architectures for
scalable quantum computing have been presented in ion-trap\cite{wineland}
and one-dimensional array of solid-state systems\cite{lidar1}.

In this letter, we show how to carry out SFTQC in DFS subblocks
for qubit systems with tunable $XXZ$ exchange interaction. By
using appropriate encoding, one can make quantum information in
every subblock interaction-free from the other subblocks even
though the physical couplings to connect different DFS subblocks
is always on. This is exactly the central idea of interaction-free
encoding\cite{zhou}. To implement nonlocal operations between
different subblocks one can transform local encoded states to
``switch on'' the interaction between adjacent subblocks. All the
operations ensure the evolution of the system in decoherence-free
subspaces. Our scheme leads to three additional prominent
advantages. First, since interaction-free encodings spontaneously
screen the subblocks from the
couplings to other subblocks it is not necessary to use recoupling pulses%
\cite{Waugh} to eliminate the effect of fixed couplings between
subblocks. Second, physical switches are not necessary for
interaction-free subspaces, and this reduces the technical
difficulties for the design of quantum computers. Finally, in our
scheme, the universal encoded QC can be realized by manipulating
only one type of interaction parameter. This further reduces the
difficulty of the manipulation.

{\bf DFS and IFS:} Let us consider a one dimensional array of
subblocks. Each subblock, which contains a certain number of
qubits, is surrounded by an independent environment. We assume
that the subblocks are only coupled to their nearest subblocks
(see Fig.1).

\begin{figure}[b]
\includegraphics[width=8cm]{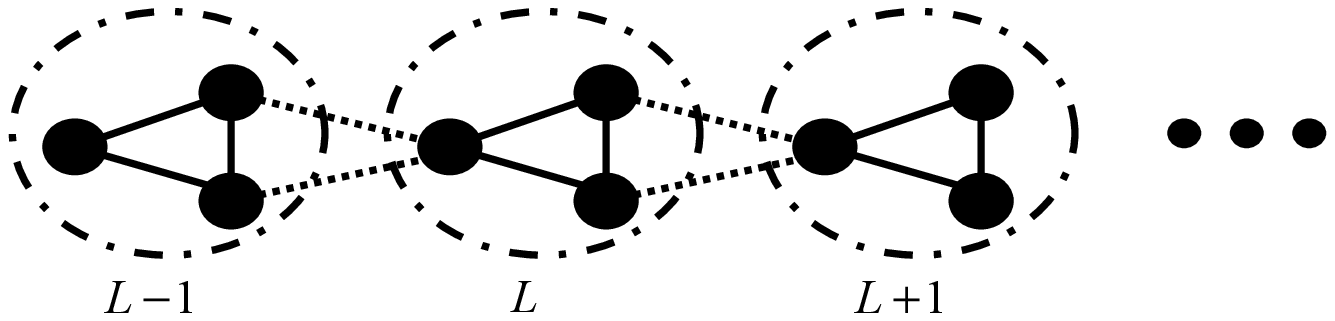}
\caption{A one dimensional array of subblocks. Each subblock
surrounded by dotted-dashed lines suffers independent environment
noise. Solid lines represent the couplings between the qubits in
one subblock. Dotted lines represent the couplings between qubits
in different subblocks.}
\end{figure}

For the $Lth$ subblock, there are three types of interactions to drive the
evolution of the qubits in it.
\begin{equation}
H_{total}=H_{LL}+H_{LB}+\sum_{L^{\prime }=L\pm 1}H_{LL^{\prime }}.
\label{eq1}
\end{equation}
Here, $H_{LL}$, $H_{LB}$ and $H_{LL^{\prime }}$ refer to interactions
between qubits in the $Lth$ subblock, qubits in the $Lth$ subblock and its
environment, and interactions between adjacent subblocks. (For simplicity,
we omit the free Hamiltonian of single qubit in this letter by using two
degenerate states or working in the rotating frame.) In general, these
interaction terms have the following forms: $H_{LL}=\sum_{%
{i<j \atop \alpha ,\beta }%
}J_{ij}^{\alpha \beta }\sigma _{Li}^\alpha \sigma _{Lj}^\beta $, $%
H_{LL^{\prime }}=\sum_{%
{i,i^{\prime } \atop \alpha ,\beta }%
}g_{ii^{\prime }}^{\alpha \beta }\sigma _{Li}^\alpha \sigma _{L^{\prime
}i^{\prime }}^\beta $, $H_{LB}=\sum_{\alpha ,i}\gamma _i^\alpha \sigma
_{Li}^\alpha b_{Li}^\alpha $. Here, $\sigma _{L\left( L^{\prime }\right)
i}^{\alpha (\beta )}$ refers to the Pauli operator of the $ith$ qubit in the
$L(L^{\prime })th$ subblock, $\alpha (\beta )=x,y,z$. $b_{Li}^\alpha $ is
the environmental operator which couples to the $ith$ qubit in the $Lth$
subblock. $J_{ij}^{\alpha \beta },g_{ii^{\prime }}^{\alpha \beta }$ and $%
\gamma _i^\alpha $ are coupling coefficients. According to our
assumption, noise from different subblocks are independent:
$\left[ b_L^\alpha ,b_{L^{\prime }}^\beta \right] =0$ for $\forall
\alpha ,\beta $ and $L\neq L^{\prime }$. Moreover, all the qubits
in each subblock encounter collective decoherence\cite{DFS}:
$\gamma _i^\alpha =\gamma ^\alpha ,b_{Li}^\alpha
=b_L^\alpha $. Thus, $H_{LB}=\sum_\alpha S^\alpha B^\alpha $, where $%
S^\alpha =\sum_i\sigma _{Li}^\alpha $, $B^\alpha =\gamma ^\alpha b_L^\alpha $%
. For the qubits in the $Lth$ subblock, due to interaction with
the environment and other subblocks the evolution is usually
nonunitary. To preserve the coherence of the qubits in the $Lth$
subblock, we wish all the couplings ($H_{LB}$ and $H_{LL^{\prime
}}$) be switched off in the idle mode. However, couplings between
the system and the environment are usually unavoidable. This
obstacle can be overcome by DFS encoding. The main idea on DFS is
that by taking advantage of the symmetry of the interaction
between
the system and the environment, one may find a special subspace ${\bf H}%
_{DFS}^L$ in the $Lth$ subblock Hilbert space ${\bf H}^L$ such that
\begin{equation}
\forall \alpha :S^\alpha \left| \psi _L^D\right\rangle =c^\alpha \left| \psi
_L^D\right\rangle ,c^\alpha \in R  \label{eq1a}
\end{equation}
if $\left| \psi _L^D\right\rangle $ $\in H_{DFS}^L$. Thus, for any initially
unentangled system-environment state $\left| \psi _L^D\right\rangle \otimes
\left| \psi _L^B\right\rangle $, $H_{LB}\left| \psi _L^D\right\rangle
\otimes \left| \psi _L^B\right\rangle =\left| \psi _L^D\right\rangle \otimes
\sum_\alpha c^\alpha B^\alpha \left| \psi _L^B\right\rangle $, where $\left|
\psi _L^B\right\rangle $ is the state of the environment. The space ${\bf H}%
_{DFS}^L$ is the so called decoherence-free subspace (DFS) in which $H_{LB}$
equivalently reduces to an environmental operator so that the effect from
the environment can be eliminated. In general, DFS can be defined by
stabilizers\cite{UFTQC1a}. In our model, since collective operators $%
S^\alpha $ are Hermitian we may introduce a type of single
parameter stabilizer $D_\Gamma ^L=\prod\limits_\alpha \exp
[-\Gamma \left( S^\alpha -c^\alpha I\right) ^2]$, which is an
identity operator on the DFS state in ${\bf H}^L$:
\begin{equation}
D_\Gamma ^L\left| \psi _L^D\right\rangle =\left| \psi _L^D\right\rangle
,iff\left| \psi _L^D\right\rangle \in {\bf H}_{DFS}^L  \label{eq1b}
\end{equation}
where $\Gamma $ is a positive real number. The stabilizer $D_\Gamma ^L$
reduces to the projector $P_{DFS}^L$ of DFS when $\Gamma \rightarrow \infty $%
:
\begin{equation}
P_{DFS}^L=\lim_{\Gamma \rightarrow \infty }D_\Gamma ^L.  \label{eq1c}
\end{equation}

Since switching the couplings between adjacent subblocks
complicates the operation and may add additional noise to the
system, we would prefer to avoid such physical switching.
Recently, there have been QC proposals with always-on
couplings\cite{zhou,Benjamin}. If the couplings between subblocks
are fixed and always on, the persistent interactions between
adjacent subblocks will distort their states in idle mode. To
eliminate this effect the idea of IFS is introduced. Assume that
the couplings between adjacent subblocks have highly symmetrical forms: $%
H_{LL^{\prime }}=\sum_\alpha g_{LL^{\prime }}^\alpha A_L^\alpha A_{L^{\prime
}}^\alpha $, where $A_{{L}\left( L^{\prime }\right) }^\alpha $ is the
collective operator of the $L(L^{\prime })th$ subblock. We may define the
interaction free subspace(IFS) ${\bf H}_{IFS}^L$ in Hilbert space ${\bf H}^L$%
, which satisfies $A_L^\alpha \left| \psi _L^I\right\rangle =a_L^\alpha
\left| \psi _L^I\right\rangle $ for any $\left| \psi _L^I\right\rangle \in
{\bf H}_{IFS}^L$. Similar to the case of DFS, the interaction Hamiltonian $%
H_{LL^{\prime }}$ reduces to an operator $\sum_\alpha g_{LL^{\prime
}}^\alpha a_L^\alpha A_{L^{\prime }}^\alpha $ acting on the $L^{\prime }th$
subblock Hilbert space ${\bf H}^{L^{\prime }}$ when the state of the $Lth$
subblock is in IFS. Thus, the stabilizer and projector of IFS of the $Lth$
subblock have the following form: $I_\Gamma ^L=\prod\limits_\alpha \exp
[-\Gamma \left( A_L^\alpha -a_L^\alpha I\right) ^2]$ and $%
P_{IFS}^L=\lim_{\Gamma \rightarrow \infty }I_\Gamma ^L$, respectively.

Since DFS and IFS separately keep the system from interacting with
the environment and other subblocks, the intersection space of DFS
and IFS naturally screen all couplings from them. For the $Lth$
subblock, we define the intersection space of DFS and IFS by ${\bf
H}_{I-D}^L$, whose projector is $P_{I-D}^L=P_{IFS}^LP_{DFS}^L$.
The dimension $d_{I-D}$ of the intersection
space ${\bf H}_{I-D}^L$ can be obtained by tracing its projector: $%
d_{I-D}=TrP_{I-D}^L$. Usually the space ${\bf H}_{I-D}^L$ is trivial, $%
d_{I-D}=0$. Only for suitably designed systems the intersection space is
nontrivial. Clearly, $d_{I-D}\geqslant 2$ is necessary for encoding quantum
information. We find that the coupling Hamiltonian $H_{LL^{\prime }}$ can
have different effects when the states in the $Lth$ and $L^{\prime }th$
subblocks are encoded in different ways. By combining the ideas of DFS and
IFS, we provide an architecture of scalable fault-tolerant quantum computer
with fixed couplings between subblocks. For a carefully designed QC system
we may have $P_{I-D}^L\neq P_{DFS}^L$ and $d_{I-D}\geqslant 2$. Thus, in
idle mode we encode quantum information in the subspace ${\bf \otimes }_L%
{\bf H}_{I-D}^L$. The local fault-tolerant QC in the $Lth$ subblock can be
realized by dynamical processes in the subspace ${\bf \otimes }_L{\bf H}%
_{I-D}^L$ or $...{\bf \otimes H}_{I-D}^{L-1}{\bf \otimes H}_{DFS}^L{\bf %
\otimes H}_{I-D}^{L+1}...$, in which the fixed couplings $H_{L,L\pm 1}$
reduce to constants or local operators. Nonlocal fault-tolerant QC between
the $Lth$ and $\left( L+1\right) th$ subblocks can be realized by the
dynamical precesses in the space $...{\bf \otimes H}_{I-D}^{L-1}{\bf \otimes
H}_{DFS}^L{\bf \otimes H}_{DFS}^{L+1}{\bf \otimes H}_{I-D}^{L+2}...$. To
preserve the state within a subspace the controllable Hamiltonian should
satisfy the following condition, which follows from theorem 3 in\cite{UFTQC2}%
.

{\em The necessary and sufficient condition for Hamiltonian H to keep the
state at all times entirely within subspace }${\bf H}_s${\em \ of Hilbert
space }${\bf H}_t${\em \ is }$\left[ P,H\right] ${\em =0, where P is the
projector of the subspace }${\bf H}_s${\em .}

In the following we will concretely present our architecture based on
tunable $XXZ$ exchange interaction.

{\bf SFTQC based on tunable }$XXZ$ {\bf interaction: }The tunable $XXZ$ type
interaction has the following form:
\begin{equation}
H_{ij}^{XXZ}=K_{ij}\left( \sigma _i^x\sigma _j^x+\sigma _i^y\sigma
_j^y\right) +J_{ij}\sigma _i^z\sigma _j^z.  \label{eq2}
\end{equation}
Here, $i$ and $j$ refer to two neighboring qubits. We assume that the
parameter $J_{ij}$ is fixed but $K_{ij}$ can be tuned. This model describes
the interactions in a few physical systems, such as electrons floating on
helium\cite{Platzman} and charge qubits based on superconducting charge boxes%
\cite{Zhouxx}. Initially, in idle mode, we adjust the coupling
coefficients $K_{ij}=0$, and $H_{ij}^{XXZ}$ reduces to the Ising
type.

We consider the case of collective phase decoherence: $H_{LB}=\sum_{i=1}^m%
\sigma _{Li}^zB_L^z$, where $m$ is the number of qubits in one subblock. DFS
of the $Lth$ subblock is the eigenspace of the operator $S_L^z=\sum_{i=1}^m%
\sigma _{Li}^z$ which has eigenvalues $l=-m,-m+2,...,m$. The dimension of
DFS with eigenvalue $l$ is $d_l=C_m^{\left( m-l\right) /2}$.

We consider subblocks composed of four particles, which are
symmetrically arranged on the vertices of a rectangle (see Fig.2).
We choose a special DFS with zero eigenvalue of $S_L^z$. This is a
6-dimensional space. For two neighboring subblocks $L$ and $L+1$,
the fixed interaction Hamiltonian has the form:
$H_{L,L+1}=J(\sigma _1^z+\sigma _2^z)\otimes (\sigma _{3^{\prime
}}^z+\sigma _{4^{\prime }}^z)$. Thus, the projectors of different
IFSs can be obtained according to the formulation in the above.
Furthermore, we may
hunt out the maximal intersection space of IFS and DFS, whose projector is $%
\sum_{i,j=0}^1\left| i_1\overline{i}_2j_3\overline{j}_4\right\rangle
\left\langle i_1\overline{i}_2j_3\overline{j}_4\right| $, where $\overline{i}%
\left( \overline{j}\right) $ refers to $1-i(j)$, $\left| 0\right\rangle $
and $\left| 1\right\rangle $ are the eigenstates of $\sigma _z$. (For
simplicity, we will arrange qubits in-order and omit their subscripts in the
following context.) This is a 4-dimensional space. we choose its
2-dimensional subspace $\{\left| 1001\right\rangle ,\left| 0101\right\rangle
\}$ as the space of logical qubit $\{\left| 0_L\right\rangle ,\left|
1_L\right\rangle \}$.

\begin{figure}[b]
\includegraphics[width=8cm]{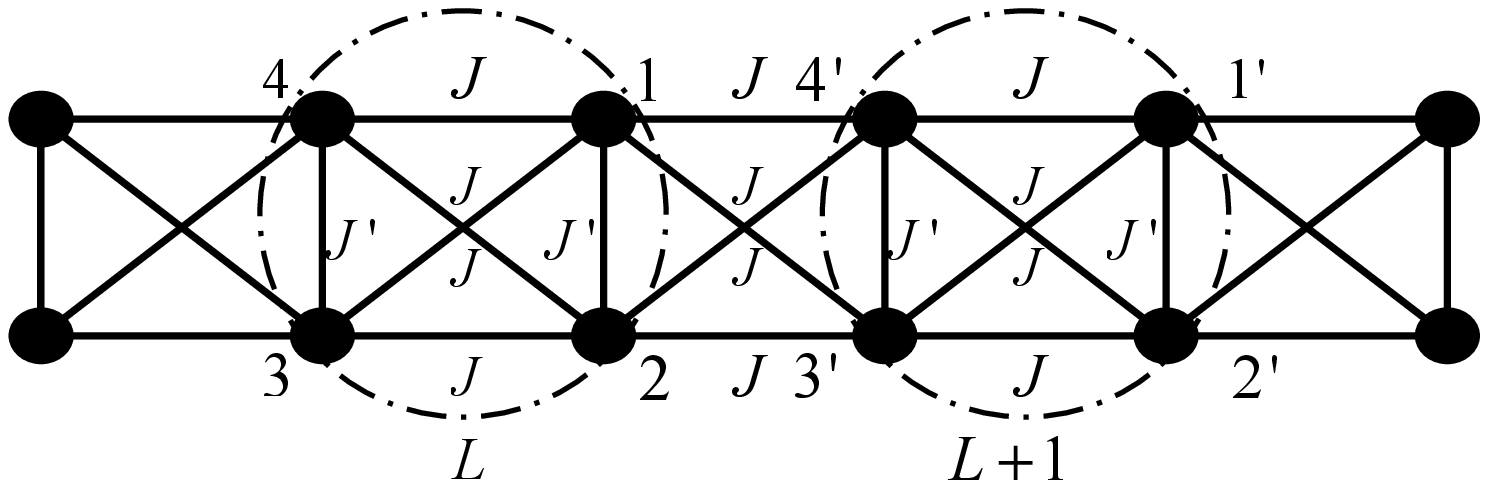}
\caption{A Strategy for concatenating different DFSs. In the idle
mode, the coupling (marked by solid lines) between qubits $i$ and
$j$ reduces to Ising type interactions ($K_{ij}=0$). The couplings
represented by horizontal and diagonal lines have the same
strength $J$. The couplings represented by vertical lines have the
strength $J^{\prime }$. Each subblock subject to local collective
dephasing environment is surrounded by dotted-dashed lines.}
\end{figure}

In our scheme for QC, we simply manipulate the interaction in
local subblocks by switching on the $K_{ij}$ coupling between some
two qubits $i$ and $j$ but fix the couplings between adjacent
subblocks.  We may prove that $\left[ H_{LL},P_{DFS}^L\right] =0 $
and $\left[ H_{L,L\pm 1},P_{DFS}^L\right] =0$. This ensures that
QC is performed in a fault-tolerant fashion.

To realize universal QC, for the $Lth$ subblock, we choose a 3-dimensional
subspace of DFS with zero eigenvalue of operator $S_L^z$ as the
computational space ${\bf H}_L^c$, which has the projector $%
P_L=\sum_{i=0}^2\left| i_L\right\rangle \left\langle i_L\right| $. Here, $%
\left| 2_L\right\rangle =\left| 0011\right\rangle _L$ is an
auxiliary state. When logical qubits are encoded in the
intersection space ${\bf H}_{I-D}^L$ spanned by $\{\left|
0_L\right\rangle ,\left| 1_L\right\rangle \}$ the effect of
nonlocal Hamiltonian coupling with the neighboring subblocks will
be eliminated. However, since the state $\left| 2_L\right\rangle $
is not in ${\bf H}_{I-D}^L$, nonlocal effects can be induced once
the quantum states in two neighboring subblocks are driven into
$\left| 2_L\right\rangle $. For simplicity, in the following we
represent the effective Hamiltonian in the
space ${\bf H}_L^c$ by $h_L$, and that in the space ${\bf H}_L^c\otimes {\bf %
H}_{L\pm 1}^c$ by $h_{L,L\pm 1}$. The operators $X_L^{lk},Y_L^{lk},$ and $%
Z_L^{lk}$ denote Pauli operators in the subspace spanned by
$\{\left| l_L\right\rangle ,\left| k_L\right\rangle \}$ $\left(
l,k=0,1,2;l\neq k\right) $.

In idle mode, $\left[ H_{LL},P_L\right] =0$, therefore the computational space ${\bf H}%
_L^c $ will be driven by the following effective Hamiltonian $h_L^{id}$:
\begin{equation}
h_L^{id}=\left[
\begin{array}{ccc}
-2J^{\prime } & 0 & 0 \\
0 & -2J^{\prime } & 0 \\
0 & 0 & 2J^{\prime }-4J
\end{array}
\right] .  \label{eq3}
\end{equation}
When we only adjust $K_{12}=\mu _L\neq 0$ ($K_{23}=\nu _L\neq 0$) the
corresponding Hamiltonian $H_{LL}^a$ ($H_{LL}^b$) satisfies $\left[
H_{LL}^{a(b)},P_L\right] =0$, which keeps the space ${\bf H}_L^c$ unchanged.
Thus, the effective Hamiltonian $h_L^a$ and $h_L^b$ have the following
forms:
\begin{eqnarray}
h_L^a &=&\left[
\begin{array}{ccc}
-2J^{\prime } & 2\mu _L & 0 \\
2\mu _L & -2J^{\prime } & 0 \\
0 & 0 & 2J^{\prime }-4J
\end{array}
\right] ,  \nonumber \\
h_L^b &=&\left[
\begin{array}{ccc}
-2J^{\prime } & 0 & 0 \\
0 & -2J^{\prime } & 2\nu _L \\
0 & 2\nu _L & 2J^{\prime }-4J
\end{array}
\right] .  \label{eq4}
\end{eqnarray}
Here, we assume that $\mu _L$ and $\nu _L$ are pre-chosen
parameters and the transition from $h_L^{id}$ to $h_L^a$ or
$h_L^b$ can be instantaneously achieved. Therefore, when we adjust
the coefficient $K_{12}$ it is equivalent to performing $X_L^{01}$
operation on the $Lth$ logical qubit. Logical $Z_L^{01}$ operation
can be implemented by switching on the $K_{23}$ coupling. After
some time $t=2\pi /\sqrt{\varsigma ^2+\left( 2\nu _L\right) ^2}$,
where $\varsigma =2\left( J-J^{\prime }\right) $, the system will
undergo an evolution given by: $U_z\left( \theta \right) =\exp
\left( -i\theta Z_L^{01}\right) $. Here, $\theta =$ $\pi \varsigma
/\sqrt{\varsigma ^2+\left( 2\nu _L\right) ^2}$. One can properly
choose the parameter $\nu _L$ to make $\varsigma /\sqrt{\varsigma
^2+\left( 2\nu _L\right) ^2}$ an irrational number. Then, positive
integer powers of $U_z\left( \theta \right) $ can approach
$U_z\left( \lambda \right) =\exp \left( -i\lambda Z_L^{01}\right)
$ to arbitrary precision, for any real $\lambda $\cite {Preskill}.
Since any single qubit gate can be decomposed into rotations
around the $z$ and $x$ axis any fault-tolerant single logical
qubit gate can be realized.

In the following we will show how to implement nonlocal operations between
the $Lth$ and $(L+1)th$ subblocks. The representation of the fixed
Hamiltonian $H_{L,L+1}$ in the space ${\bf H}_L^c\otimes {\bf H}_{L+1}^c$
has the effective form:
\begin{equation}
h_{L,L+1}=-4J\left[
\begin{array}{ccc}
0 & 0 & 0 \\
0 & 0 & 0 \\
0 & 0 & 1
\end{array}
\right] \otimes \left[
\begin{array}{ccc}
0 & 0 & 0 \\
0 & 0 & 0 \\
0 & 0 & 1
\end{array}
\right] .  \label{eq5}
\end{equation}
We find that controlled phase gates can be realized by switching on and off
the $K_{23}$ and $K_{2^{\prime }3^{\prime }}$ couplings in adjacent
subblocks $L$ and $L+1$. In this process, the whole effective Hamiltonian $%
h_{tot}$ is:
\begin{equation}
h_{tot}=h_L+h_{L+1}+h_{L,L+1}.  \label{eq6}
\end{equation}
In the following we show the implementation of controlled phase gate in two
cases.

i) $\left| J\right| ,\left| J^{\prime }\right| \ll \left| \nu _L\right|
\left( =\left| \nu _{L+1}\right| \right) $. In this case, simultaneously
switch on $K_{23}$ and $K_{2^{\prime }3^{\prime }}$ couplings, after a
suitable time interval $t=\frac \pi {4\left| \nu _L\right| }$ switch them
off. The system is driven by strong pulses and the state $\left|
1_L\right\rangle \left( \left| 1_{L+1}\right\rangle \right) $ can be flipped
to $\left| 2_L\right\rangle \left( \left| 2_{L+1}\right\rangle \right) $. In
the space $\{\{\left| 1_L\right\rangle ,\left| 2_L\right\rangle \}\otimes
\{\left| 1_{L+1}\right\rangle ,\left| 2_{L+1}\right\rangle \}\}$ the whole
idle Hamiltonian is equivalent to $%
h_{tot}^{id}=h_L^{id}+h_{L+1}^{id}+h_{L,L+1}=-5J+(3J-2J^{\prime
})\left( Z_L^{12}+Z_{L+1}^{12}\right) -JZ_L^{12}\otimes
Z_{L+1}^{12}$. Let the system evolve for a suitable time, drive it
back into IFS by using strong pulses again. A controlled phase
gate can be realized by the above 2-qubit interaction
$h_{tot}^{id}$ and local operations\cite{Chuang}. Unfortunately,
the condition i) is not realistic in certain physical systems\cite
{Platzman,Zhouxx}.

ii) We assume that $\nu _L$ and $\nu _{L+1}$ can be slowly changed
with time so that Hamiltonian $h_{tot}$ can be adiabatically
varied. To implement the nonlocal operation, we first transform
logical state $\left| 1_L\right\rangle $ to $\left|
2_L\right\rangle $. Since $h_L^{id}$ and $h_L^b $ separately
reduce to noncommuting operators $-2J+(2J-2J^{\prime })Z_L^{12}$
and $-2J+(2J-2J^{\prime })Z_L^{12}+2\nu X_L^{12}$ in the space spanned by $%
\{\left| 1_L\right\rangle ,\left| 2_L\right\rangle \}$ this transformation
can always be achieved by using Hamiltonian $h_L^{id}$ and $h_L^b$\cite
{Preskill}. After this, we switch on the $K_{2^{\prime }3^{\prime }}$
coupling in the $\left( L+1\right) th$ subblock. At that time, Hamiltonian $%
h_{L,L+1}$ starts to work. If we further assume $\left|
J-J^{\prime }\right| ,\left| 2J-J^{\prime }\right| \gg \left| \nu
_{L+1}(t)\right| $, the adiabatic condition can be reduced to
$\left| J-J^{\prime }\right|
^2,\left| 2J-J^{\prime }\right| ^2\gg \frac 18 \left| \stackrel{\cdot }{\nu }%
_{L+1}(t)\right| $. These instantaneous eigenstates of Hamiltonian $%
h_{tot}(t)=h_L^{id}+h_{L+1}^b+h_{L,L+1}$ will slowly change along
with the varying $h_{tot}(t)$. Finally, the strength of the
$K_{2^{\prime }3^{\prime }}$ coupling goes back to its starting
value $\nu _{L+1}(t_f)=\nu _{L+1}\left( 0\right) =0$, and the
instantaneous eigenstates return to their initial
forms\cite{Dittrich}. During the adiabatic evolution every logical
state will obtain corresponding phases. Since it is easy to prove
that the Berry phase is zero during this evolution we need only
calculate the dynamical phase $\varphi ^d$:
\begin{equation}
\varphi ^d=-\int_0^{t_f}\left\langle \psi \left( t\right) \right|
h_{tot}(t)\left| \psi \left( t\right) \right\rangle dt.  \label{eq7}
\end{equation}
Thus, we can obtain the corresponding phases for all combinations
of the states of the two logical qubits:
\begin{equation}
\begin{array}{l}
\varphi _{00}^d=4J^{\prime }t_f;\varphi _{01}^d\doteq 4J^{\prime }t_f+\eta
\\
\varphi _{20}^d=4Jt_f;\varphi _{21}^d\doteq 4Jt_f+\kappa
\end{array}
\label{eq8}
\end{equation}
where $\eta =\int_0^{t_f}\frac{\left( \nu _{L+1}\left( t\right) \right) ^2}{%
J^{\prime }-J}dt$ ,$\kappa =\int_0^{t_f}\frac{\left( \nu _{L+1}\left(
t\right) \right) ^2}{\left( J^{\prime }-2J\right) }dt$. This evolution can
be represented by a unitary operator $U_{eq}$ in the space $\{\{\left|
0_L\right\rangle ,\left| 2_L\right\rangle \}\otimes \{\left|
0_{L+1}\right\rangle ,\left| 1_{L+1}\right\rangle \}\}$:
\begin{equation}
U_{eq}=\exp \{i\left( a+bZ_L^{02}+cZ_{L+1}^{01}+dZ_L^{02}\otimes
Z_{L+1}^{01}\right) \}.  \label{eq9}
\end{equation}
Here, $a$, $b$, $c$ and $d$ are determined by dynamical phases
$\varphi _{mn}^d\left( m=0,2;n=0,1\right) $ and the coefficient
$d=(\kappa -\eta )/4$ is vital for the nonlocal operation. When we
appropriately adjust the function $\nu _{L+1}\left( t\right) $ and
the time of the adiabatic evolution $t_f$ we can always make $d$ a
suitable value $\theta $. Thus, the adiabatic evolution produces
an equivalent nonlocal unitary transformation in the logical
2-qubit space: $U_{eq}\equiv \exp (i\theta Z_L^{02}\otimes
Z_{L+1}^{01})$, where $\theta =\frac 14\int_0^{t_f}\frac{J\left(
\nu _{L+1}\left( t\right) \right) ^2}{\left( J^{\prime }-2J\right)
\left( J^{\prime }-J\right) }dt$. Finally, we map $\left|
2_L\right\rangle $ back to $\left| 1_L\right\rangle $. In the
evolution, the extra single logical operation produced by the
adiabatic evolution and mapping can be compensated by local
evolutions in an interaction-free fashion. Thus we have
implemented SFTQC. This architecture can be realized in certain
physical systems such as arrays of superconducting charge
qubits\cite{Makhlin} or electrons on liquid helium\cite{Platzman}.

Z-W. Zhou, Bo Yu, and G-C. Guo are funded by NFRP( 2001CB309300), NNSF of
China under Grants No.10204020. Work of X. Zhou and M. J. Feldman was
supported in part by AFOSR and funded under the DoD DURINT program and by
the ARDA.


\begin{references}
\bibitem{DFS} P. Zanardi and M. Rasetti, Phys. Rev. Lett. {\bf 79}, 3306
(1997); L.-M. Duan and G.-C. Guo, {\it ibid}. {\bf 79}, 1953
(1997).

\bibitem{DFSsys} E. Knill {\it et al.}, Phys. Rev. Lett. {\bf 84}, 2525
(2000).

\bibitem{UFTQC1} D.A. Lidar {\it et al.}, Phys. Rev. Lett. {\bf 81}, 2594
(1998).

\bibitem{UFTQC1a} D. Bacon {\it et al.}, Phys. Rev. Lett. {\bf 85}, 1758
(2000).

\bibitem{UFTQC2} J. Kempe {\it et al.}, Phys. Rev. A {\bf 63}, 042307
(2001).

\bibitem{wineland} D. Kielpinski {\it et al.}, Nature {\bf 417}, 709 (2002).

\bibitem{lidar1} D.A. Lidar and L.-A. Wu, Phys. Rev. Lett. {\bf 88},
017905 (2002); L.-A. Wu and D.A. Lidar, {\it ibid.} {\bf 88},
207902 (2002).

\bibitem{zhou} X. Zhou {\it et al.}, Phys. Rev. Lett. {\bf 89}, 197903
(2002).

\bibitem{Waugh} J.S. Waugh {\it et al.}, Phys. Rev. Lett. {\bf 20}, 180
(1968).

\bibitem{Benjamin} S.C. Benjamin and S. Bose, Phys. Rev. Lett. {\bf 90},
247901 (2003).

\bibitem{Platzman} P.M. Platzman and M.I. Dykman, Science {\bf 284}, 1967
(1999).

\bibitem{Zhouxx} X. Zhou {\it et al.}, Phys. Rev. A {\bf 69}, 030301(R)
(2004).

\bibitem{Preskill} J. Preskill, Lecture Notes for Physics 229: {\it Quantum
Information and Computation}, Caltech, 1998.

\bibitem{Chuang}  M.A. Nielsen and I.L. Chuang, {\it Quantum Computation
and Quantum Information}, Cambridge University Press, 2000.

\bibitem{Dittrich} W. Dittrich and M. Reuter, {\it Classical and Quantum
Dynamics}, Springer-Verlag, 1994.

\bibitem{Makhlin}  Y. Makhlin {\it et al.}, Rev. Mod. Phys. {\bf 73}, 357
(2001).
\end{references}
\end{document}